# ZERODUR® based optical systems for quantum gas experiments in space

**Moritz Mihm[a]\*, Jean Pierre Marburger[a], André Wenzlawski[a], Ortwin Hellmig[f], Oliver Anton[b], Klaus Döringshoff[b], Markus Krutzik[b], Achim Peters[b], Patrick Windpassinger[a], and the MAIUS Team[a,b,c,d,e,f]**

[a] *Institute of Physics, Johannes Gutenberg University Mainz, Staudingerweg 7, 55128 Mainz, Germany*, mmihm@uni-mainz.de, jemarbur@uni-mainz.de, awenzlaw@uni-mainz.de, windpass@uni-mainz.de
[b] *Department of Physics, Humboldt University of Berlin, Newtonstraße 15, 12489 Berlin, Germany*, oliver.anton@physik.hu-berlin.de, klaus.doeringshoff@physik.hu-berlin.de, markus.krutzik@physik.hu-berlin.de, achim.peters@physik.hu-berlin.de
[c] *Institute of Quantum Optics, University of Hanover, Welfengarten 1, 30167 Hannover, Germany*
[d] *Ferdinand-Braun-Institut, Leibniz-Institut für Höchstfrequenztechnik, Gustav-Kirchhoff-Str. 4, 12489 Berlin, Germany*
[e] *The Center of Applied Space Technology and Microgravity (ZARM), University of Bremen, Am Fallturm 2, 28359 Bremen, Germany*
[f] *Institute of Laser Physics, Universität Hamburg, Luruper Chaussee 149, 22761 Hamburg, Germany*, ohellmig@physnet.uni-hamburg.de
\* Corresponding Author

## Abstract

Numerous quantum technologies make use of a microgravity environment e.g. in space. Operating in this extreme environment makes high demands on the experiment and especially the laser system regarding miniaturization and power consumption as well as mechanical and thermal stability. In our systems, optical modules consisting of ZERODUR® based optical benches with free-space optics are combined with fiber components. Suitability of the technology has been demonstrated in the successful sounding rocket missions FOKUS, KALEXUS and MAIUS-1. Here, we report on our toolkit for stable optical benches including mounts, fixed and adjustable mirrors as well as polarization maintaining fiber collimators and couplers made from ZERODUR®. As an example, we present the optical modules for the scientific rocket payload of MAIUS-2, a quantum gas experiment performing dual-species atom interferometry with Bose-Einstein condensates. The modules are used on the one hand to stabilize the laser frequencies and on the other hand to distribute, overlap and switch the laser beams. This includes the overlap and joint fiber coupling of beams at 767 nm and 780 nm in the same polarization state to cool and manipulate atoms of both species simultaneously. Future projects include the development of a platform for experiments with cold atoms onboard the International Space Station. The laser system again involves ZERODUR® based optical benches in conjunction with fiber optical components. The experiment is planned as multi-user facility and currently in the design phase. The next step is to build the training, test and flight hardware.

**Keywords:** Microgravity, optical bench, laser system, integrated optics, glass ceramic, ZERODUR®

## 1. Introduction

Numerous applications of quantum technologies benefit from a microgravity environment. The operation under rough conditions makes high demands on the experiment and especially the laser system regarding miniaturization and power consumption as well as mechanical and thermal stability. We have developed a technology for stable optical benches to overcome these difficulties. Our optical modules are based on ZERODUR®, a glass ceramic providing high mechanical and thermal stability with a thermal expansion coefficient as low as $\pm 0.007 \cdot 10^{-6}$/K in the range between 0°C and 50°C.

Here, we report on our toolkit for stable optical benches including mounts, fixed and adjustable mirrors as well as polarization maintaining fiber collimators and couplers made from ZERODUR®. Systems based on our toolkit can fulfil all sorts of functions including laser frequency stabilization, switching and distribution of laser light. Fast switching is achieved by acousto-optic modulators (AOMs), perfect extinction by integrated mechanical shutters. As the newest tool, we have implemented an overlap and joint fiber coupling of beams at 767 nm and 780 nm in the same polarization state.

Our technology has been developed in the context of atom interferometry in space. Like numerous quantum technologies, atom interferometry benefits from a microgravity environment. On ground, the time of the matter wave in the interferometer is naturally limited by the free-fall time of the atoms. Longer interrogation times can be achieved on various microgravity platforms providing different quality and duration of microgravity. Common platforms include parabolic flights [1], drop towers [2,3], sounding rockets [4], the International Space Station and satellites. Suitability of our technology has been demonstrated in the successful sounding rocket missions MAIUS-1 [5], KALEXUS [6] and FOKUS [7].

As an exemplary application of our technology, we present the optical modules for the scientific rocket payload of MAIUS-2, a quantum gas experiment performing dual-species atom interferometry with Bose-Einstein condensates. Future projects include the development of the multi-user facility BECCAL, a platform for experiments with cold atoms aboard the International Space Station. From the experiences on sounding rockets, ZERODUR® based optical benches are again the technology of choice for the laser system.

After introducing the environmental and experimental requirements, we discuss our jointing techniques and toolkit for stable ZERODUR® based optical benches. Finally, the optical modules for the rocket payload of MAIUS-2 are presented.

## 2. Requirements

Quantum optical experiments usually require the possibility to manipulate, overlap, distribute and switch laser light in free space with low losses. Guiding the light to and from the modules is typically realized by optical fibers. One of the main challenges when working with fibers is the coupling of light into the fiber. Due to the small mode field diameters, positioning accuracy in the sub-micrometer range and high stability is required. We use polarization-maintaining single-mode fibers (Nufern PM780-HP). In order to minimize system losses, the non-angled end-faces have antireflective coatings.

During launch phase, our sounding rocket payloads undergo typical accelerations up to 13 g and vibrations up to $1.8\,g_{RMS}$. We perform environmental tests imposing vibrations both on component and system level with loads of $5.4\,g_{RMS}$ and $8.1\,g_{RMS}$, respectively [5].

## 3. Jointing technique

The assembly of stable optical benches requires jointing techniques providing stable bonds and allowing for precise alignment of the components. Common techniques include hydroxide-catalysis bonding and adhesive bonding.

The hydroxide-catalysis bonding technique was patented at Stanford University [8,9]. It is performed in cleanroom environments between flat polished surfaces with an overall global flatness of at least λ/10. Once applied, the alignment of components must be done in approximately 120 s and therefore requires an alignment jig. On the other hand, achieving full bonding strength takes four weeks at ambient temperature [10].

Adhesive bonding using e.g. the two-component epoxy Hysol EA 9313 from Henkel can be realized on matte surfaces in a laboratory environment [11]. Handling strength is achieved after 8 h (1 h with external heating), full strength after five days at ambient temperature.

Our technology takes advantage of the long processing times (several hours) of two light-curing adhesives while curing can be achieved within one minute once radiated at the appropriate wavelengths [12]. The ZERODUR® itself is optically transparent at these wavelengths allowing to build components sequentially i.e. curing outer parts of complex components without affecting inner parts. In this way, we assemble and integrate components before aligning and locking inner parts in place. For example, we can align the polarization and beam pointing by adjusting the fiber inside the collimator even after the housing of the fiber collimator has already been glued onto the baseplate. To glue optical elements on ZERODUR® we use Norland 63 that cures at wavelengths around 365 nm. For gluing ZERODUR® on ZERODUR® we use Fusion Flo, a nanocomposite that cures at wavelengths around 460 nm.

As third adhesive, we apply Crystalbond 509 to components designated to be realigned after being glued on the optical bench (waveplates/adjustable mirrors). Crystalbond 509 is solid at room temperature and gets viscous when heated (softening point at 71°C). This is a reversible process allowing to conveniently realign components whenever necessary.

When gluing Non- ZERODUR® components on ZERODUR®, the mismatch in thermal expansion has to be considered. Whereas the surface area/shape of Non-ZERODUR® components changes with temperature, the coefficient of thermal expansion of ZERODUR® is negligible. This can lead to a weakened connection or even cause components to break down. The mismatch applies in particular to active components and is reinforced by the fact that ZERODUR® is a poor thermal conductor. We use the two-component epoxy Hysol Tra-Bond F112 which can endure thermal stress to mount active components like e.g. AOMs but also to lock optical isolators into position.

## 4. Toolkit

Fig. 1 gives an overview of our toolkit for stable optical benches including fiber collimators, optical isolators, AOMs, rotation mounts, beam splitting cubes, fixed and adjustable mirrors and fiber couplers.

Baseline for the assembly of an optical module is to glue all free-space components sequentially (from fiber collimator to coupler) on the ZERODUR® baseplate. Consequently, the following component can compensate for misalignments of the previous component. Exceptions are components with predefined positions such as AOMs and optical isolators as well as the very last components in the beam paths (fiber couplers). The fiber coupler is specifically designed to compensate for all misalignments on the way whereas our fiber collimator is designed to define a fix starting point for beam paths. To account for misalignments when securing the fiber ferrule inside a fiber coupler, an adjustable mirror is placed right in front.

The design of our fixed and adjustable mirrors, beam splitting cubes, rotation mounts, fiber collimators and couplers has already been introduced in [12].

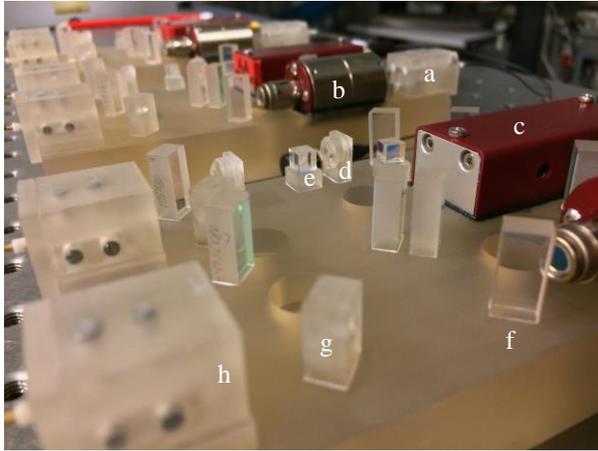

Fig. 1. Picture of a ZERODUR® based optical module showing our toolkit of components including fiber collimator (a), optical isolator (b), AOM (c), rotation mount (d), beam splitting cube (e), fixed (f) and adjustable mirror (g) and fiber coupler (h).

To integrate free-space optical isolators (Medium Power Faraday Isolator from Electro-Optics Technology, Inc.) and AOMs (MT80-A1.5-IR from AA Sa), cut-outs are milled into the baseplates to maintain the beam height on the modules. In the case of the cylindrical isolators, the cut-outs are also cylindrical, while the cut-outs for the AOMs are flat. Both the optical isolators and the AOMs are fixed with Fusion Flo in the first place before filling purposive gaps underneath with Hysol Tra-Bond F112. This adhesive can endure the thermal stress produced by the AOMs. Prior to integration, components are characterized. Results of the characterization of ten AOMs are an average first order diffraction efficiency at 1.3 W RF power (reference value of the manufacturer) of $(95 \pm 1)$% and an average rise time (10% to 90%; 0.95 mm $1/e^2$ beam diameter) of $(149.5 \pm 1.3)$ ns. Due to our mounting concept we have all degrees of freedom to reproduce these values on the optical modules.

On the modules for the payload of MAIUS-2 (see Sec. 5), we have implemented the overlap and joint fiber coupling of beams at 767 nm and 780 nm. Furthermore, to allow for polarization cleaning behind the optical fibers, all beams are in the same polarization state. We use dichroic mirrors to overlap the beams of different wavelengths. The coatings are applied on polished ZERODUR® substrates. When gluing the components with Fusion Flo on the baseplate, the dichroic mirror is aligned for minimal reflection of the transmitted beam at 780 nm. Losses below 3% are achieved. With the dichroic mirror locked into position, two adjustable mirrors are glued on the baseplate in the beam path of light at 767 nm to overlap beams of both wavelengths (see Fig. 2). It is crucial that these two mirrors are glued with centrally aligned beams to prevent displacements when realigning. Afterwards, beam walks are performed with the adjustable mirrors to achieve the best overlap possible. To couple the light of both wavelengths into an optical fiber, the fiber ferrule position is aligned radially with light at 780 nm. Due to the chromatic aberration of the coupling lens (Thorlabs 355230-B), the focal difference at 767 nm and 780 nm is approximately 4 μm and the fiber end-face is aligned laterally in-between the two foci. After securing the fiber ferrule, the beam walk is repeated to maximize the coupling efficiency at 767 nm. In practice, average coupling efficiencies of 86% at both wavelengths are achieved, where we define the efficiency as the ratio of power in the optical fiber to the power in front of the fiber coupler. The chromatic shift causes a theoretical loss of efficiency of approximately 1% while we observe a loss in the range of 3%-5% in practice. The fibers have an anti-reflective coating to reduce reflections at the fiber end-faces.

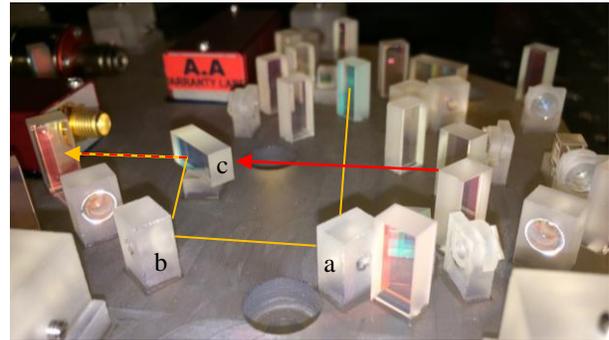

Fig. 2. Picture of a configuration to overlap beams at the wavelengths 767 nm (yellow) and 780 nm (red) including two adjustable mirrors (a, b) and a dichroic mirror (c).

## 5. Optical benches

MAIUS-2 is a quantum gas experiment performing dual-species atom interferometry with Bose-Einstein condensates on sounding rockets. The light provided by the laser system is guided via optical fibers to the physics package for 2D/3D-MOT and interferometry with Bose-Einstein condensates of potassium and rubidium atoms. MAIUS-2 is the follow-up to the MAIUS-1 experiment which successfully created the first Bose-Einstein condensate in space during a sounding rocket flight on the 23rd of January 2017 [4].

The laser system of MAIUS-2 contains in total ten diode lasers on a water-cooled platform. Five lasers at 767 nm are dedicated to transitions of potassium atoms, four lasers at 780 nm to transitions of rubidium atoms. Another laser at 1064 nm is used for a dipole trap. With the exception of the dipole laser, all lasers are fiber connected to overall seven ZERODUR® based optical modules. For each of the two species, one module provides an atomic frequency reference. Two lasers are frequency stabilized on the atomic transitions, seven science lasers are frequency stabilized w.r.t. the two reference lasers via offset beat notes. The science lasers in turn are connected to three ZERODUR® based single-input modules and two double-input modules to distribute, overlap and switch the laser beams. The layout of these modules is further discussed in the following sections.

### 5.1 Single-input module

A picture of one of the three single-input modules is shown in Fig. 3. Each of the three modules has one input and two output ports. Optical isolators are integrated behind the fiber collimators to prevent back reflections. Zeroth and first diffraction order of the AOMs are separated and the first order is split for the two output ports while the zeroth order is used for intensity monitoring. The intensity ratio between the two output ports can be adjusted using a waveplate and polarizing beam splitter cube. Two modules are dedicated to potassium and one to rubidium.

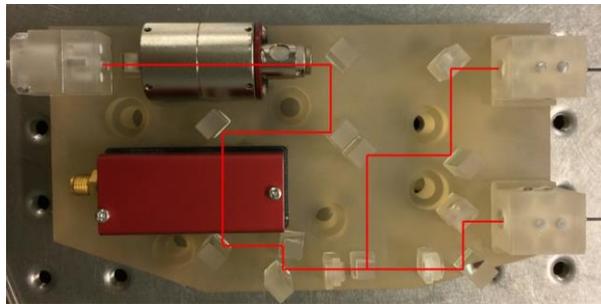

Fig. 3. Picture of one of the three ZERODUR® based single-input modules for the scientific rocket payload of MAIUS-2.

### 5.2 Double-input module

One of the two double-input modules is depicted in Fig. 4. At each of the two input ports, one laser dedicated to transitions of potassium and rubidium atoms is connected. The beam paths of light at the respective wavelengths 767 nm (yellow) and 780 nm (red) are shown. Behind both fiber collimators, optical isolators are integrated to prevent back reflections and

AOMs to switch the beams. Following the beam path of light at 780 nm, light of the first diffraction order is split off (output 4 and 5) with adjustable intensity using a waveplate and polarizing beam splitter cube. Afterwards the beam is overlapped (indicated by dashed lines) with light at 767 nm at a dichroic mirror and jointly fiber coupled (output 1 and 2). The zeroth diffraction order is directly overlapped and fiber coupled with light at 767 nm (output 3).

## 6. Conclusion

We introduced our toolkit for stable optical modules consisting of ZERODUR® based optical benches with free-space optics. Providing high mechanical and thermal stability, the technology is ideally suited for laser systems in space or other field applications.

As an example of use, the layout and purpose of the optical modules for the scientific rocket payload MAIUS-2 have been presented. The modules are currently characterized and prepared for integration into the complete laser system. A picture of one of the two ZERODUR® based frequency reference modules in the laser system housing during an integration test and frequency stability measurement is shown in Fig. 5. The next steps include the completion of characterization tests, final integration and the linkage with the other experiment subsystems.

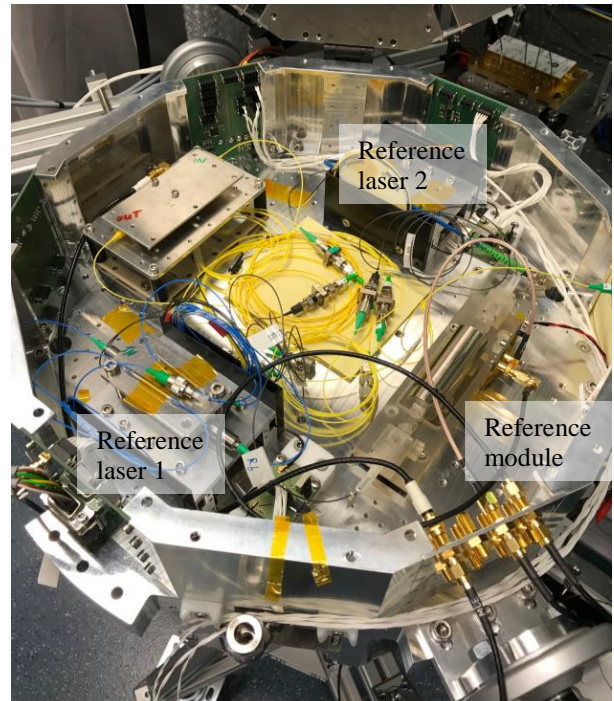

Fig. 5. Picture of one of the two ZERODUR® based frequency reference modules (right) in the laser system housing during integration test and frequency stability measurement.

## Acknowledgements

Our work is supported by the German Space Agency DLR with funds provided by the Federal Ministry for Economic Affairs and Energy (BMWi) under grant numbers 50 WP 1433 and 50 WP 1703.

Special thanks to Klaus Sengstock from the University of Hamburg for supporting the development of the ZERODUR® Technology.

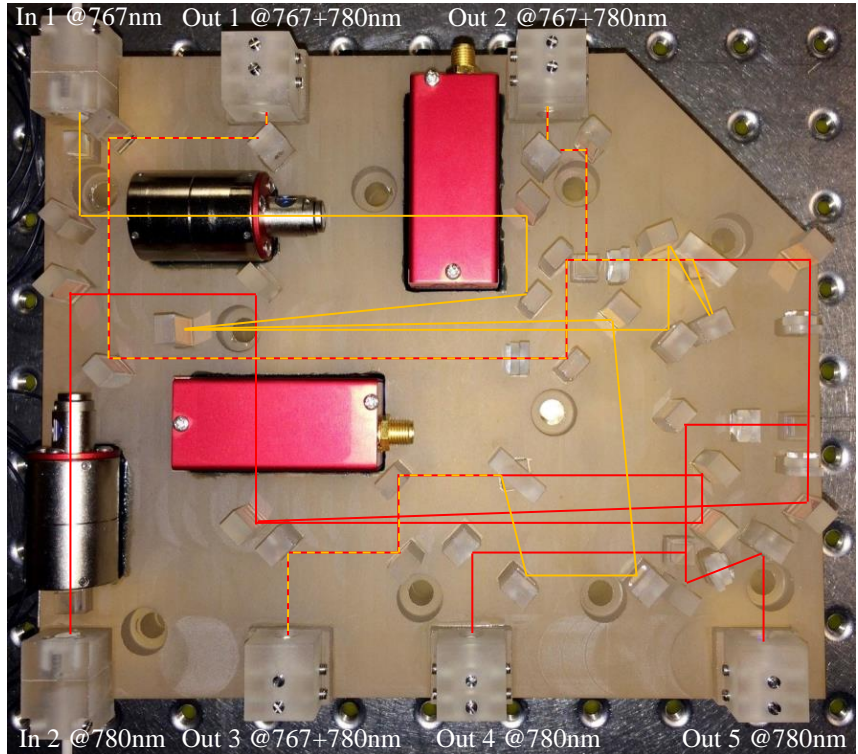

Fig. 4. Picture of one of the two ZERODUR® based double-input modules for the scientific rocket payload of MAIUS-2. The beam paths of light at the wavelengths 767 nm (yellow) and 780 nm (red) are shown. After splitting light at 780 nm off (output 4 and 5), beams are overlapped (dashed) with light at 767 nm at dichroic mirrors and jointly fiber coupled (output 1 to 3).

## References

[1] B. Barrett, L. Antoni-Micollier, L. Chichet, B. Battelier, T. Lévèque, A. Landragin, P. Bouyer, Dual matter-wave inertial sensors in weightlessness, Nature communications 7 (2016).

[2] T. van Zoest, N. Gaaloul, Y. Singh, H. Ahlers, W. Herr, S.T. Seidel, W. Ertmer, E. Rasel, M. Eckart, E. Kajari, S. Arnold, G. Nandi, W.P. Schleich, R. Walser, A. Vogel, K. Sengstock, K. Bongs, W. Lewoczko-Adamczyk, M. Schiemangk, T. Schuldt, A. Peters, T. Könemann, H. Müntinga, C. Lämmerzahl, H. Dittus, T. Steinmetz, T.W. Hänsch, J. Reichel, Bose-Einstein condensation in microgravity, Science 328 (2010) 1540-1543.

[3] H. Müntinga, H. Ahlers, M. Krutzik, A. Wenzlawski, S. Arnold, D. Becker, K. Bongs, H. Dittus, H. Duncker, N. Gaaloul, C. Gherasim, E. Giese, C. Grzeschik, T.W. Hänsch, O. Hellmig, W. Herr, S. Herrmann, E. Kajari, S. Kleinert, C. Lämmerzahl, W. Lewoczko-Adamczyk, J. Malcolm, N. Meyer, R. Nolte, A. Peters M. Popp, J. Reichel, A. Roura, J. Rudolph, M. Schiemangk, M. Schneider, S.T. Seidel, K. Sengstock, V. Tamma, T. Valenzuela, A. Vogel, R. Walser, T. Wendrich, P. Windpassinger, W. Zeller, T. van Zoest, W. Ertmer, W. P. Schleich, and E.M. Rasel, Interferometry with Bose-Einstein condensates in microgravity, Phys. Rev. Lett. 110 (2013).

[4] M. D. Lachmann, H. Ahlers, D. Becker, S.T. Seidel, T. Wendrich, E. M. Rasel, W. Ertmer, Creating the first Bose-Einstein condensate in space, Proc. of SPIE Vol. 10549 (2018).

[5] V. Schkolnik, O. Hellmig, A. Wenzlawski, J. Grosse, A. Kohfeldt, K. Döringshoff, A. Wicht, P. Windpassinger, K. Sengstock, C. Braxmaier, M. Krutzik, A. Peters, A compact and robust diode laser system for atom interferometry on a sounding rocket, Appl. Phys. B 122:217 (2016).

[6] A.N. Dinkelaker, M. Schiemangk, V. Schkolnik, A. Kenyon, K. Lampmann, A. Wenzlawski, P. Windpassinger, O. Hellmig, T. Wendrich, E.M. Rasel, M. Giunta, C. Deutsch, C. Kürbis, R. Smol, A. Wicht, M. Krutzik, A. Peters, Autonomous frequency stabilization of two extended-cavity diode lasers at the potassium wavelength on a sounding rocket, Applied Optics 56 (2017) 1388-1396.

[7] M. Lezius, T. Wilken, C. Deutsch, M. Giunta, O. Mandel, A. Thaller, V. Schkolnik, M. Schiemangk, A. Dinkelaker, A. Kohfeldt, A. Wicht, M. Krutzik, A. Peters, O. Hellmig, H. Duncker, K. Sengstock, P. Windpassinger, K. Lampmann, T. Hülsing, T.W. Hänsch, R. Holzwarth, Space-borne frequency comb metrology, Optica 3 (2016) 1381-1387.

[8] D.-H. Gwo, Ultra precision and reliable bonding method, US Patent 6,284,085 B1 (2001).

[9] D.-H. Gwo, Hydroxide-catalyzed bonding, US Patent 6,548,176 B1 (2003).

[10] S. Ressel, M. Gohlke, D. Rauen, T. Schuldt, W. Kronast, U. Mescheder, U. Johann, D. Weise, C. Braxmaier, Ultrastable assembly and integration technology for ground- and space-based optical systems, Applied Optics 49 (2010) 4296-4303.

[11] M. Gohlke, T. Schuldt, K. Döringshoff, A. Peters, U. Johann, D. Weise, C. Braxmaier, Adhesive Bonding for Optical Metrology Systems in Space Applications, J. Phys.: Conf. Ser. 610 (2015).

[12] H. Duncker, O. Hellmig, A. Wenzlawski, A. Grote, A.J. Rafipoor, M. Rafipoor, K. Sengstock, P. Windpassinger, Ultrastable, Zerodur-based optical benches for quantum gas experiments, Applied Optics 53 (2014) 4468-4474.